\documentclass[11pt]{article}
\usepackage[letterpaper]{geometry}
\usepackage[parfill]{parskip}
\usepackage{amsmath,amsthm,amssymb,bbm}
\usepackage{mathtools}
\usepackage{cases}
\usepackage{graphicx}
\usepackage{tabularx}
\usepackage{microtype}
\usepackage{enumitem}

\usepackage{authblk}

\usepackage{url}
\usepackage[colorlinks,citecolor=blue,urlcolor=blue,linkcolor=blue,linktocpage=true]{hyperref}
\usepackage{cleveref}
\crefformat{equation}{(#2#1#3)}
\crefrangeformat{equation}{(#3#1#4) to~(#5#2#6)}
\crefname{equation}{}{}
\Crefname{equation}{}{}

\usepackage[authoryear]{natbib}
\usepackage{multirow}

\usepackage{algpseudocode,algorithm}

\newtheoremstyle{mythmstyle}
  {8 pt} 
  {3 pt} 
  {} 
  {} 
  {\bfseries} 
  {.} 
  {.5em} 
  {} 

\theoremstyle{plain}

\renewcommand{\hat}{\widehat}

\makeatletter
\def\thm@space@setup{%
  \thm@preskip=6pt plus 1pt minus 1pt
  \thm@postskip=\thm@preskip 
}
\makeatother

\newtheorem{theorem}{Theorem}[section]
\newtheorem{lemma}[theorem]{Lemma}
\newtheorem{corollary}[theorem]{Corollary}

\newtheorem{remark}{Remark}

\newtheorem*{example*}{Example}
\newtheorem{definition}{Definition}
\newtheorem*{definition*}{Definition}
\newtheorem*{remark*}{Remark}
\crefname{definition}{\textbf{definition}}{definitions}
\Crefname{definition}{Definition}{Definitions}
\crefname{assumption}{\textbf{assumption}}{assumptions}
\Crefname{assumption}{Assumption}{Assumptions}

\newcommand{\myscale}{0.6}

\begin{document}
\allowdisplaybreaks
\title{Fast Exact Conformalization of Lasso using Piecewise Linear Homotopy}

 \author[1]{Jing Lei}
\affil[1]{Department of Statistics, Carnegie Mellon University}

\maketitle

\begin{abstract}
Conformal prediction is a general method that converts almost any point
predictor to a prediction set.  The resulting set
keeps good statistical properties of the original estimator under standard
assumptions, and guarantees valid average coverage even when the model
is misspecified.  A main challenge in applying conformal prediction in modern
applications is efficient computation, as it generally requires an exhaustive search
over the entire output space.  In this paper we develop an exact and computationally
efficient conformalization of the Lasso and elastic net.  The method makes use
of a novel piecewise linear homotopy of the Lasso solution under perturbation of a single input sample point.
As a by-product, we provide a simpler and better justified online Lasso algorithm,
which may be of independent interest.
Our derivation also reveals an interesting accuracy-stability trade-off in conformal inference,
which is analogous to the bias-variance trade-off in traditional parameter estimation.
The practical performance of the new algorithm is demonstrated using
both synthetic and real data examples.
\end{abstract}


\section{Introduction}
\label{sec:introduction}


Conformal prediction is a generic technique that converts point estimators, such
as density estimators, regression function estimators, and cluster estimators,
to prediction sets with valid coverage under very weak assumptions.  Conformal
prediction features an ``out-of-sample'' fitting approach, which re-fits
the model by augmenting the data set with a candidate data point.  Based on the
re-fitting, it calculates a conformity score for the candidate data point,
which measures how well this data point agrees with the sample and the model.
Under a symmetry assumption of the fitting procedure and exchangeability of
sample points, the conformity score can be transformed to a valid $p$-value.
The prediction set is then obtained by thresholding the $p$-values obtained at
all candidate data points.

Conformal prediction has two attractive properties. First and most importantly, 
conformal prediction offers \emph{model-free}
coverage guarantee.  On average, conformal prediction sets always cover
at least the nominal level as long as the data points are exchangeable and the
fitting procedure is symmetric with respect to the input data points, even when the
model is completely misspecified. Second, conformal prediction can be combined
with almost any existing point estimators. Even if an estimator is asymmetric,
one can always construct a symmetric version using a U-statistic type transform.

Since its first appearance as an online learning tool \citep{VovkGS05,vovk2009line},
conformal prediction has been developed in the statistical literature in a few
directions. Including optimal choices of conformity score and statistical
efficiency \citep{LeiRW13,LeiW14}, efficient computation for complex input data
types or fitting procedures \citep{Hebiri10,LeiRW14,chen2016trimmed},
and classification \citep{Lei14,sadinle2016least}.
Recently, \cite{LeiGRTW16} systematically extended the conformal prediction
method to high dimensional regression. 

Despite the attractive properties and all the progress, 
the application of conformal prediction often comes at a high computational cost,
especially for
high-dimensional complex data.
By definition, in order to compute the conformal prediction set, one needs to
exhaustively search all points in the sample space, where for each point
the model needs to be re-fitted and the conformity score needs to be re-calculated.
Therefore, the form of a conformal prediction set depends on the initial estimator
as well as the conformity score.  For some simple cases, close-form characterizations of
conformal prediction sets are available, such as least
squares regression or ridge regression \citep{burnaev2014efficiency}.  Efficient
approximations are available for kernel density estimator \citep{LeiRW13} and
kernel nonparametric regression \citep{LeiW14}.

In high-dimensional problems, the estimators are inevitably more complicated and
hence the corresponding conformal prediction sets are much harder to
characterize.  On the other hand, conformal prediction is most useful in high
dimensional settings as the model assumptions such as sparsity and low-intrinsic
dimensionality are often not true, and the inference tools developed under such
assumptions are often invalid under model misspecification.

Our starting point of high-dimensional conformal prediction is the Lasso \citep{lasso},
one of the most popular and well-studied
sparse linear regression methods. To compute the conformal
prediction set at a new covariate vector $x$, the algorithm implemented
in the \texttt{conformalInference} package developed in \cite{LeiGRTW16} scans
a set of grid points in the $y$-space, 
each with a new Lasso fitting and ranking of the fitted residuals. The output
is essentially an evaluation of the indicator function of the prediction set
at these grid points. If the grid size is 100, then this procedure
costs more than 100 times computing resource as the original Lasso fit, 
for just a single value of
covariate vector $x$, which is rather prohibitive in practice.

There are several attempts to efficiently conformalize the Lasso.  \cite{LeiGRTW16}
considered a sample splitting technique developed in  \cite{LeiRW14} 
that detaches the fitting step and ranking
step (see also
\cite{VovkGS05}, under the name of \emph{inductive conformal prediction}).
This method loses some statistical efficiency, as both the
fitting and ranking are carried out on a reduced sample size. Moreover,
the sample splitting adds another layer of randomness which may be undesirable
for the construction of prediction intervals.
\cite{Hebiri10} proposed a partial conformalization of Lasso that uses the fitted
residuals in a similar way as the original conformal prediction.  However, that algorithm
is not fully symmetric regarding the augmented sample and hence
loses the key property of model free coverage guarantee.
More recently, \cite{chen2016trimmed} proposed a method of finding a smaller
search space.  This method, which may reduce the computational cost by a
constant fraction, is still based on evaluating the indicator function of the 
prediction set on a set of grid points.

In this paper, we develop an algorithm that efficiently and exactly computes the
conformal prediction set for the Lasso.  Unlike the grid search based methods,
we prove that the Lasso conformal prediction sets are unions of intervals and
our algorithm finds the exact end points of these intervals from a single Lasso fitting
on the original input data set.
Our main technique is a novel
piecewise linear homotopy of Lasso solution under perturbation of a single sample
point of the training data.  Such a piecewise linear homotopy allows us to explicitly 
track the re-fitted residuals as a piecewise linear function of the varying
candidate value of $y$, which in turn leads to a simple rule of determining if
$y$ is in the conformal prediction set.

In addition to the main contribution, our derivation leads to the following interesting
observations.
\begin{enumerate}
  \item Due to the piecewise linearity of the refitted residuals, the conformal
  prediction set of Lasso is a union of intervals.  In most cases, the prediction
  set is a single interval.  We provide sufficient conditions for the prediction
  set to be an interval.  Such conditions can be easily verified in extensions
  such as the elastic net.
  \item Our derivation also reveals an interesting accuracy-stability trade-off
  in conformal inference, which is analogous to the well-known bias-variance trade-off.
  Here the term ``accuracy'' refers to the magnitude of fitted residuals, and
  the term ``stability'' refers to how fast the fitted residuals change when
  one response value changes.  When the estimator is heavily regularized, the fit is more
  stable but less accurate, and vice versa.  The most statistically efficient 
  (i.e., the shortest) conformal
  prediction set is achieved when the two competing criteria
  are balanced.
  \item Our perturb-one analysis of Lasso can also be used to efficiently update the Lasso solution
  when the data points arrive sequentially.  A non-linear homotopy algorithm for
   online Lasso has been developed by \cite{garrigues2009homotopy}. Our algorithm
   is piecewise linear and hence easier to implement.  Moreover, \cite{garrigues2009homotopy}
   only derives formulas for finding the points of change on the homotopy path, it does
   not provide validity guarantee of continuation after
   the points of change.  In contrast, our derivation provides
   a more complete theoretical justification, including validity of continuation
   under a natural assumption that each point of change involves only a single
   coordinate.
\end{enumerate}

In \Cref{sec:prelim} we briefly introduce conformal prediction and its combination
with the Lasso.  In \Cref{sec:method} we derive the piecewise linear homotopy algorithm
with theoretical justification, and investigate sufficient conditions under which the prediction set
is an interval.  In \Cref{sec:data} we demonstrate the performance of the algorithm
on synthetic and real data examples.  In \Cref{sec:discussion} we conclude the
paper with some discussion.

\section{Background}
\label{sec:prelim}

\subsection{Conformal prediction}
Given independent and identically distributed data $(Z_i)_{i=1}^n$ in $\mathbb R^d$
from an underlying distribution $P$, the problem of prediction set is to find
a $\hat C\subseteq \mathbb R^d$ such that $P(Z_{n+1}\in \hat C)\ge 1-\alpha$,
where $\alpha \in(0,1)$ is a prescribed error level.

This setup includes both unsupervised and supervised statistical learning.
For example, in unsupervised learning problems such as density estimation or 
clustering, $Z_i$'s are the data points and there is no response variable.
In supervised learning problems such as linear regression, $Z_i=(X_i,Y_i)$
is the covariate-response pair, such that $X_i\in\mathbb R^p$, $Y_i\in\mathbb R^1$
and $p=d-1$.  The prediction set $\hat C$ gives a prediction set of $Y$ for each
$X$ by defining $\hat C(x)=\{y\in\mathbb R^1: (x,y)\in\hat C\}$.

The general recipe of conformal prediction starts from a function 
$$A(z_1,...,z_n;z_{n+1}): \mathbb R^{d\times(n+1)}\mapsto \mathbb R^1\,,$$
such that $A$ is symmetric in the first $n$ inputs.  In the following
we define $\vec z=(z_1,...,z_{n+1})$ and
 $\vec z_{-i}$ by removing $z_i$ in $\vec z$ for $i=1,...,n+1$.

The function $A$ is called the \emph{conformity score} function.  It has two
components.  The first is a modeling component, in that $A$ usually involves
a model fitting step.  The second is a deviation (or conformity) component, in that 
$A$ also measures how well the last input agrees with the fitted model.
For example, in density estimation, we can define $A(\vec z_{-(n+1)}; z_{n+1})=\hat f(z_{n+1})$,
where $\hat f$ is an estimated density function of $P$ using the augmented data set $\vec z$.
In regression, we can define $A(\vec z_{-(n+1)};z_{n+1})=-|y_{n+1}-\hat f(x_{n+1})|$
where $\hat f$ is an estimated regression function using data $\vec z$.
A higher value of conformity implies better agreement between the sample point
and the fitted model.

Given an $x_{n+1}$, at which a prediction of $Y$ is wanted, 
the conformal prediction method proceeds by  defining a $p$-value for each candidate
value $y\in\mathbb R^1$:
\begin{align}
  \hat p_y =1- \frac{1}{n+1}\sum_{i=1}^{n+1}\mathbf 1\left(A_i \ge A_{n+1}\right)
\end{align}
where $A_i=A(\vec z_{-i};z_i)$ for $1\le i\le n+1$  and $y_{n+1}=y$.
The conformal prediction set is then obtained by thresholding the $p$-values
at $\alpha$:
\begin{align}\label{eq:hat_C}
  \hat C(x_{n+1})=\{y: \hat p_y\ge \alpha\}\,.
\end{align}

It can be shown \citep{LeiRW13,LeiGRTW16,VovkGS05} that under the assumption that
$(X_i,Y_i)_{i=1}^{n+1}$ are exchangeable,
then $\hat p_{Y_{n+1}}$ has a sub-uniform distribution and is hence a valid $p$-value.
Therefore we have
\begin{equation}\label{eq:coverage}
P(Y_{n+1}\in \hat C(X_{n+1}))\ge 1-\alpha\,.
\end{equation}
Such a coverage guarantee requires only exchangeability of input data and
symmetry of the conformity score function
$A$, which holds for almost all popular model fitting algorithms. The probabilistic
statement about the coverage needs to be interpreted with care, as detailed in
the following remark.

\begin{remark}
  The probability in \eqref{eq:coverage} refers to the joint distribution of
  $(X_i,Y_i)_{i=1}^{n+1}$.  Therefore, the coverage guarantee is in an average
  sense.  It averages over the training sample $(X_i,Y_i)_{i=1}^n$, as well
  as the new data pair $(X_{n+1},Y_{n+1})$.  In other words, the actual coverage
  may be more than $1-\alpha$ for some training sample or some values $X_{n+1}$,
  and less than $1-\alpha$  for other combinations of training sample and $X_{n+1}$.
  This shall \emph{not} be interpreted as a weakness of conformal prediction.
  In fact, one can also show \citep{LeiW14,LeiGRTW16} that when
  the initial model estimator is accurate, which typically holds under standard
  regularity conditions, then the conformal prediction set
  is nearly optimal and may provide asymptotic conditional coverage
  $$
  P(Y\in \hat C(x)\mid X = x)= 1-\alpha+o_P(1)\,~~\forall x\,.
  $$
  So conformal prediction shall be regarded as an additional
  level of protection against potential model misspecification, producing
  valid average coverage even when the regularity conditions are violated
  and the fitting is arbitrarily bad. 
\end{remark}

\subsection{Conformal prediction with Lasso}
Now we focus on linear regression where $Z_i=(X_i,Y_i)$ with $X_i\in \mathbb R^p$,
$Y_i\in\mathbb R^1$, and $p$ can be much larger than $n$.
Suppose we are given a paired training sample of size $n$, $(x_i, y_i)_{i=1}^n$,
and would like to construct a conformal prediction set of $Y$ at 
future covariate $X_{n+1}=x_{n+1}$
using the Lasso as the base estimator.
Following the general recipe described above, we only need to specify the
conformity score function $A$, which in turn requires specification
of the model fitting method and the conformity measure.


For any estimated regression coefficient vector $\hat\beta$, it is natural
to use the fitted residual to measure the ``agreement'' or ``conformity''
of a data point $(x,y)$:
\begin{equation}\label{eq:conformity}
  -|y-x'\hat\beta|\,.
\end{equation}

Let $\|\cdot\|_1$ denote the vector $\ell_1$ norm.
 For a given tuning parameter $\lambda >0$, we consider the Lasso estimate
 using the augmented data
\begin{equation}\label{eq:lasso_aug}
  \tilde\beta(y)=\arg\min_{\beta\in\mathbb R^p}\frac{1}{2}\sum_{i=1}^n(y_i-x_i'\beta)^2
  +\frac{1}{2}(y-x_{n+1}'\beta)^2
  +\lambda\|\beta\|_1\,.
\end{equation}
That is, $\tilde\beta(y)$ denotes the
Lasso estimator one would obtain when including $(x_{n+1},y)$ as the $(n+1)$th
data point. 
Here $y_{n+1}=y$ is a candidate value in $\mathbb R^1$.
Using the notation of the general recipe and the convention
$y_{n+1}=y$, we have, for a given tuning parameter
$\lambda>0$,
\begin{align*}
  A_i(y) = -|y_i - x_i'\tilde\beta(y)|,~~i=1,...,n+1\,.
\end{align*}

According to the definition of the conformal prediction set $\hat C(x)$,
$y\in \hat C(x_{n+1})$ if and only if
$A_{n+1}(y)$ ranks no higher than $(n+1)\alpha$ among $\{A_{i}(y):1\le i\le n+1\}$ in ascending order.  A brute-force method of computing
$\hat C(x_{n+1})$ would be to calculate and rank $A_{i}(y)$ for all $y$ and all $i$,
which is practically infeasible.  In practice one can only evaluate $\mathbf 1_{\hat C(x_{n+1})}(y)$ over a fine grid of $y$ values.

\section{Efficient and exact conformalization of Lasso}
\label{sec:method}
\subsection{A piecewise linear homotopy for the perturb-one Lasso}
Now we derive our main technical component of our method: a piecewise linear
homotopy of the Lasso solution. 
To simplify notation and facilitate discussion, we use the following notation
and terminologies.
\begin{itemize}
  \item  $\hat\beta$ is
the Lasso solution using the original data set $(x_i,y_i)_{i=1}^n$.
\item $y_{n+1}(t)=x_{n+1}'\hat\beta+t$ for a fixed $x_{n+1}\in\mathbb R^p$. 
Our discussion will focus on positive
values of $t$ but the derivation extends easily to the negative values as well.
\item 
We use $\hat\beta(t)$ to denote the Lasso solution one would get by
adding $(x_{n+1},y_{n+1}(t))$ as the $(n+1)$th sample point.
\item $v(t)$ denotes the dual variable of the Lasso problem, see \Cref{eq:opt_condition_beta_t} below.
\item For a matrix $\Sigma$ and a set $J$ of indices, $\Sigma_J$ denotes the diagonal submatrix indexed by $J$.
For a vector $x$, $x_{J}$ denotes the subvector indexed by $J$.
\item $\hat\Sigma=n^{-1}\sum_{i=1}^n x_i x_i'$ is the sample covariance matrix. 
\end{itemize}
We will show that, under appropriate conditions,
starting from $t_0=0$, we can find an increasing sequence $(t_k:k=0,1,2,...)$
such that
\begin{enumerate}
  \item the support of $\hat\beta(t)$ is
constant, denoted by $J_k$, on each interval $[t_k,t_{k+1})$;
\item $\hat\beta_{J_k}(t)$ is a linear function of $t$ on $[t_k,t_{k+1}]$,
with a slope vector denoted by $\eta(k)\in\mathbb R^{|J_k|}$;
\item $v_{J_k^c}(t)$ is a linear function of $t$ on $[t_{k},t_{k+1}]$ with a slope
vector denoted by $\gamma(k)\in\mathbb R^{|J_k^c|}$.
\end{enumerate}

%

In the following we derive the exact formulas for all the objects of interest 
in a recursive manner, including $t_k$, $J_k$, $\eta(k)$, and $\gamma(k)$.

\subsubsection{The initial piece of the homotopy}\label{sec:initial}
  Using optimality condition of $\hat\beta$ we have
  \begin{equation}\label{eq:opt_condition_beta}
    -\sum_{i=1}^n(y_i-x_i'\hat\beta)x_i +  v =0\,,
  \end{equation}
  where $v\in \mathbb R^p$ is the dual variable satisfying $v_j={\rm sign}(\hat\beta_j)\lambda$ 
  if $\hat\beta_j\neq 0$ and $v_j\in[-\lambda,\lambda]$ if $\hat\beta_j=0$.

When $t=0$, we have $y_{n+1}(0)=x_{n+1}'\hat\beta$ so that \eqref{eq:opt_condition_beta} can be written as
\begin{equation}\label{eq:optimality_beta_0}
  -\sum_{i=1}^n(y_i-x_i'\hat\beta)x_i-(y_{n+1}(0)-x_{n+1}'\hat\beta)x_{n+1} +  v =0\,,
\end{equation}
which implies the following result.
\begin{lemma}\label{lem:beta_0}
  $\hat\beta(0)=\hat\beta\,.$
\end{lemma}

Next we investigate properties of $\hat\beta(t)$ for $t\ge 0$.  The discussion
and derivation for $t\le 0$ is analogous and omitted.
In general, the optimality condition of $\hat\beta(t)$ can be written as
\begin{equation}\label{eq:opt_condition_beta_t} -\sum_{i=1}^n(y_i-x_i'\hat\beta(t))x_i-(y_{n+1}(t)-x_{n+1}'\hat\beta(t))x_{n+1}+ v(t)=0\,,
\end{equation}
where $v(t)\in\mathbb R^p$ is the corresponding dual variable.


We start from $k=0$ and $t_0=0$. By definition, $J_0=\{j:\hat\beta_j(0)\neq 0\}$. 
Now assume that the dual variables $v_j$ ($j\in J_0^c$) are not on the boundary:
$\max_{j\in J_0^c} |v_j|<\lambda$. Then for small values of $t$ 
we shall expect $\hat\beta(t)$ to have the same support with the same signs 
as $\hat\beta(0)$.
In particular, we can write the optimality condition for $\hat\beta(t)$, assuming it
has the same signs as $\hat\beta(0)$.
\begin{align}
  &-\sum_{i=1}^n(y_i-x_{i}'\hat\beta(t))x_{i,J_0}-(y_{n+1}(t)-x_{n+1}'\hat\beta(t))x_{n+1,J_0}+ v_{J_0}=0\,,\label{eq:optimality_beta_t_1}\\
  &\|v_{J_0^c}(t)\|_\infty\le \lambda\,,\label{eq:optimality_beta_t_2}
\end{align}
where 
\begin{equation}\label{eq:v_t}
v_{J_0^c}(t)  =\sum_{i=1}^n(y_i-x_{i}'\hat\beta(t))x_{i,J_0^c}+(y_{n+1}(t)-x_{n+1}'\hat\beta(t))x_{n+1,J_0^c}
\end{equation}
Combining \eqref{eq:optimality_beta_0} and \eqref{eq:optimality_beta_t_1} we have
\begin{equation}\label{eq:update_rule}
  \hat\beta_{J_0}(t)=\hat\beta_{J_0}(0)+\eta(0) t\,,
\end{equation}
where
\begin{equation}\label{eq:eta}
  \eta(0)=\left(\sum_{i=1}^{n+1} x_{i,J_0} x_{i,J_0}'\right)^{-1}x_{n+1,J_0} = \frac{n^{-1}\hat\Sigma_{J_0}^{-1}x_{n+1,J_0}}{1+n^{-1}x_{n+1,J_0}'\hat\Sigma_{J_0}^{-1}x_{n+1,J_0}}\,.
\end{equation}
The last equality in \eqref{eq:eta} uses the Sherman-Morrison identity.

For the coordinates in $J_0^c$, we can combine \eqref{eq:optimality_beta_0} with
\eqref{eq:v_t} and \eqref{eq:update_rule}, which leads to
\begin{equation}\label{eq:v_t_homotopy}
  v_{J_0^c}(t) = v_{J_0^c}(0)+\gamma(0) t
\end{equation}
where
\begin{align}
  \gamma(0)=&x_{n+1,J_0^c}-\left(\sum_{i=1}^{n+1} x_{i,J_0^c}x_{i,J_0}'\right)
  \left(\sum_{i=1}^{n+1}x_{i,J_0}x_{i,J_0}\right)^{-1}x_{n+1,J_0}\nonumber\\
  =&\frac{x_{n+1,J_0^c}-\hat\Sigma_{J_0^c,J_0}\hat\Sigma_{J_0}^{-1}x_{n+1,J_0}}{1+n^{-1}x_{n+1,J_0}\hat\Sigma_{J_0}^{-1}x_{n+1,J_0}}\,.\label{eq:gamma}
\end{align}
In order for $\hat\beta(t)$ given in \eqref{eq:update_rule} to be a valid
solution, it is sufficient and necessary to satisfy the sign constraint:
\begin{equation}\label{eq:sign_constraint}
{\rm sign}(\hat\beta_{J_0}(t))={\rm sign}(\hat\beta_{J_0}(0))\,  
\end{equation}
and the dual variable bound
\begin{equation}\label{eq:v_t_condition}
\left\|v_{J_0^c}(t)\right\|_\infty=  \left\|  v_{J_0^c}(0) + \gamma(0) t \right\|_\infty\le \lambda\,.
\end{equation}
  Therefore, the largest
positive value $t$ that satisfies both constraints are
\begin{equation}\label{eq:t_1}
  t_1\coloneqq \min_{j\in J_0}\left(-\frac{\hat\beta_j(0)}{\eta_j(0)}\right)_{++} \bigwedge
  \min_{j\in J_0^c}\left(\frac{{\rm sign}(\gamma_j(0))\lambda-v_j(0)}{\gamma_j(0)}\right)_{++} 
\end{equation}
where for a real number $z$, $(z)_{++}$ equals $z$ if $z>0$ and equals $\infty$
if $z\le 0$.  In \eqref{eq:t_1}, we use the convention that $0/0=0$, $z/0={\rm sign}(z)\infty$ if $z\neq 0$.

Now we have completed the first piece in the piecewise linear homotopy. 
Next we show how to find directions in the following pieces of the homotopy.

\subsubsection{The points of change and validity of continuation}

\begin{definition}
We call $t\in\mathbb R$ a \emph{point of change}
  if the solution $\hat\beta(t)$ of
  the augmented the Lasso problem \eqref{eq:lasso_aug} with $y_{n+1}=x_{n+1}'\hat\beta+t$
   changes its support at $t$.  
\end{definition}
By definition, $t_1$ given in \eqref{eq:t_1} is such a point of change, 
since for $t$ slightly larger than $t_1$
the support of $\hat\beta(t)$ will be different from $J_0$.
Suppose that the support
changes to $J_1$.  Then one would expect to be able to define $\eta(1)$ 
similarly as in \eqref{eq:eta} by replacing
$J_0$ with $J_1$, with a similar rule of update as in
\eqref{eq:update_rule} applicable to $t\in[t_1,t_2]$ for some $t_2>t_1$.

Formally, \Cref{thm:continuation} below establishes
a recursive way to find all points of change as well as explicit formulas for
both the primal and dual variables as functions of $t$, under the
assumption that the support of $\hat\beta(t)$ changes
in only one coordinate at each point of change.  Similar formulas can be derived for negative values of $t$.
\begin{theorem}\label{thm:continuation}
  Let $t_0=0$, $J_0=\{j:\hat\beta(0)\neq 0\}$, and for $k\ge 0$ define 
    \begin{align}
    \eta(k)&=\frac{n^{-1}\hat\Sigma_{J_{k}}^{-1}x_{n+1,J_{k}}}{1+n^{-1}x_{n+1,J_{k}}
    \hat\Sigma_{J_{k}}^{-1}x_{n+1,J_{k}}}\,,\label{eq:rec_eta+1}\\
    \gamma(k)&=\frac{x_{n+1,J_{k}^c}-\hat\Sigma_{J_{k-1}^c,J_{k}}
    \hat\Sigma_{J_{k}}^{-1}x_{n+1,J_{k}}}{1+n^{-1}x_{n+1,J_{k}}
    \hat\Sigma_{J_{k}}^{-1}x_{n+1,J_{k}}}\,.\label{eq:rec_gamma+1}\\
    \hat\beta_{J_{k}^c}(t)&=0~~~\text{and}~~~\hat\beta_{J_{k}}(t)=
    \hat\beta_{J_{k}}(t_{k})+\eta(k)(t-t_k)\,,~~\forall~t\in[t_k,t_{k+1}]\,,\label{eq:rec_primal+1}\\
    v_{J_{k}^c}(t)&=v_{J_{k}^c}(t_{k}) +\gamma(k) (t-t_{k})\,,~~\forall~t\in[t_k,t_{k+1}]\,,
    \label{eq:rec_dual+1}\\
    t_{k+1}
    &=t_{k}+\min_{j\in J_{k}}\left(-\frac{\hat\beta_j(t_{k})}{\eta_j(k)}\right)_{++} \bigwedge
    \min_{j\in J_{k}^c}\left(\frac{{\rm sign}(\gamma_j(k))\lambda-v_j(t_{k})}{\gamma_j(k)}\right)_{++}\,,\label{eq:t_k+1}\\
    J_{k+1}&=\left\{\begin{array}{ll}
    J\backslash\{j\} & \text{if the minimum of \eqref{eq:t_k+1} is achieved by $j\in J_{k}$,}\\
    J\cup\{j\} & \text{if the minimum of \eqref{eq:t_k+1} is achieved by $j\in J_{k}^c$.}\end{array}
    \right.\label{eq:rec_J_k}
  \end{align}
  Assume $\|v_{J_0^c}(0)\|_{\infty}<\lambda$ and for some $K\ge 1$,
  \begin{enumerate}
    \item $\hat\Sigma_{J_{k}}$ has full rank for $0\le k\le K-1$,
    \item \eqref{eq:t_k+1} achieves its minimum at only one $j$ for $1\le k\le K-1$.
  \end{enumerate}
  Then for all $t\in[0,t_{K}]$, 
  $\hat\beta(t)$ is the unique Lasso solution corresponding to the data set $(x_i,y_i)_{i=1}^{n+1}$
  with $y_{n+1}=y_{n+1}(t)$ and tuning parameter $\lambda$.
\end{theorem}


In the recursion, the case of $K=0$ has been treated in \Cref{sec:initial}. The remaining of the proof
amounts to verify that the primal variable given in \eqref{eq:rec_primal+1} is
a valid Lasso solution for $t\in [t_{k},t_{k+1}]$, provided that the result holds
for $t\in [0,t_{k}]$, $k=1,...,K-1$. We call this the \emph{validity of continuation}.
Here we cannot directly repeat the argument
used for the initial piece because by construction of $t_k$
the assumption $\|v_{J_k^c}(t)\|_\infty < \lambda$ is violated at $t=t_k$.

\Cref{thm:continuation} uses a simple and useful sufficient condition for validity
of continuation by assuming that only one coordinate of inactive variables has its dual variable on
the boundary when the support changes.
We call this the \emph{simple change assumption}, formally given below.
\begin{definition} We say a point of change $t=t_k$ ($k=1,2,...,$) is \emph{simple} if only one coordinate
  achieves the minimum in \eqref{eq:t_k+1}.
\end{definition}

A few remarks are in order before we prove \Cref{thm:continuation}.

\begin{remark}\label{rem:validity}
  A nonlinear piecewise smooth homotopy of the Lasso after adding a single data point has been
  proposed by \cite{garrigues2009homotopy}. That paper aims at finding
  the Lasso solution after adding the $(n+1)$th data point $(x_{n+1},y_{n+1})$,
  by considering adding $(tx_{n+1},ty_{n+1})$ as the $(n+1)$th data point and
  letting $t$ change from $0$ to $1$.  It derives formulas for finding
  the points of change, but does not prove validity of continuation.  
  In this paper, \Cref{thm:continuation} 
  provides a simpler and more theoretically justifiable solution
  to the same problem, since one can use the piecewise linear homotopy
  by varying $t$ from $0$ to $y_{n+1}-x_{n+1}'\hat\beta$.
  Our piecewise linear homotopy makes use of the observation that
  adding data point $(x_{n+1}, x_{n+1}'\hat\beta)$ does not change the solution
  (\Cref{lem:beta_0}). The linearity of the primal and dual updates also makes it
  easier to check validity of continuation.  Finally, we note that
  the homotopy considered in \cite{garrigues2009homotopy} cannot be used
  to derive the conformal prediction set of Lasso.
\end{remark}

\begin{remark}\label{rem:simple}
  The requirement of the change points being simple is to make the calculation
  more tractable.  When the joint distribution of $(X,Y)$ is continuous,
  a point of change being non-simple implies that two jointly continuous variables are 
  equal, which has probability 0.  This assumption has never
  been violated in our numerical experiments.  In the rare case of a non-simple point of
  change, the homotopy is still piecewise linear, but finding the explicit formulas
 for the active set and the slopes becomes more complicated.
\end{remark}

\begin{remark}\label{rem:full_rank}
  The requirement of $\hat\Sigma_{J_k}$ being full-rank is also natural.
  When it is not full rank, the Lasso solution is not unique, which happens
  with zero probability if the columns of design matrix are generally positioned
  \citep{Tibshirani13}.  In the rare case that $\hat\Sigma_{J_k}$ is rank deficient,
  one can add a small ridge penalty term to avoid the problem.  We will
  extend our results to the case of elastic net in \Cref{sec:en} below.
\end{remark}

\begin{proof}[Proof of \Cref{thm:continuation}]
  The initial piece with $k=0$ has been established in \Cref{sec:initial}. 

The case of $K=1$ requires no more proof.  When $K\ge 2$, we prove by induction. 
  Assume that the claim holds for $t\in [t_{k},t_{k+1}]$ for some $0\le k\le K-2$.
  It suffices to prove validity of continuation for $t\in [t_{k+1},t_{k+2}]$.
  
  Consider two cases.
  \begin{enumerate}
    \item [] Case 1. At $t_{k+1}$, one active variable in $J_{k}$ becomes inactive. This corresponds to
    the scenario that
    the minimum of \eqref{eq:t_k+1} is achieved at some $j\in J_{k}$.
    \item [] Case 2. At $t_{k+1}$, one inactive variable in $J_{k}^c$ becomes active.
    This corresponds to the scenario that the minimum of \eqref{eq:t_k+1}
    is achieved at some $j\in J_{k}^c$.
  \end{enumerate}
  Under the assumption that $t_{k+1}$ is a simple point of change, these two
  disjoint cases cover all possibilities.
  
  In Case 1, let $j\in J_{k}$ be the coordinate that $\hat\beta_{j}(t_{k+1})=0$.
We know that $\eta_j(k)$ and $v_j(t_{k+1})$ must have different signs, because
$0=\hat\beta_{j}(t_{k+1})=\hat\beta_{j}(t_{k})+\eta_j(k)(t_{k+1}-t_{k})$ and
$v_j(t)={\rm sign}(\hat\beta_{j}(t_{k}))\lambda$ for all $t\in[t_{k},t_{k+1}]$.
The conjectured dual variable is, for $\epsilon>0$ small enough,
\begin{equation}
  v_j(t_{k+1}+\epsilon) = v_j(t_{k+1})+\gamma_j(k+1)\epsilon=
  {\rm sign}(\hat\beta_{j}(t_{k}))\lambda+\gamma_j(k+1)\epsilon\,.
\end{equation} As a result,
in order to establish validity of continuation, it suffices to show that
$\gamma_j(k+1)$ and $\eta_j(k)$ have the same sign.

Using  \eqref{eq:rec_eta+1} we have
\begin{align*}
  \eta(k) =& c\hat\Sigma_{J_{k}}^{-1}x_{n+1,J_{k}}
\end{align*}
for some $c>0$.
Using the fact that $J_{k}=J_k\cup \{j\}$ and blockwise matrix inversion,
we have
\begin{align*}
  \eta_j(k) = c\left(\hat\Sigma_{jj}-\hat\Sigma_{j,J_{k+1}}\hat\Sigma_{J_{k+1}}^{-1}\hat\Sigma_{J_{k+1},j}\right)^{-1}
  \left(x_{n+1,j}-\hat\Sigma_{j,J_{k+1}}\hat\Sigma_{J_{k+1}}^{-1}x_{n+1,J_{k+1}}\right)
\end{align*}
On the other hand, using
\eqref{eq:rec_gamma+1}  have
\begin{align*}
  \gamma_j(k+1)=&c'(x_{n+1,j}-\hat\Sigma_{j,J_{k+1}}\hat\Sigma_{J_{k+1}}^{-1} x_{n+1,J_{k+1}})
\end{align*}
for some $c'>0$.
Therefore we confirm that $\gamma_{j}(k+1)$ and $\eta_j(k)$ have the same sign.

In Case 2, let $j\in J_{k}^c$ be such that $v_j(t_{k+1})\in\{\pm \lambda\}$.  By construction $\hat\beta_{j}(t_{k+1})=0$
and the conjectured solution $\hat\beta_j(t_{k+1}+\epsilon)=\eta_j(k+1)\epsilon$ for $\epsilon>0$ small enough.
In order to establish validity of continuation, we need to show that $\eta_j(k+1)$
has the same sign as $v_j(t_{k+1})$.

Using \eqref{eq:rec_eta+1} we know that $\eta_j(k+1)$
is the $j$-coordinate of $c\hat\Sigma_{J_{k+1}}^{-1}x_{n+1,J_{k+1}}$
for some $c>0$. Using blockwise inversion we have
\begin{align*}
  \eta_j(k+1)=&c\left(\hat\Sigma_{jj}-
  \hat\Sigma_{j,J_{k}}\hat\Sigma_{J_{k}}^{-1}\hat\Sigma_{J_{k},j}\right)^{-1}
  \left(x_{n+1,j}-\hat\Sigma_{j,J_{k}}\hat\Sigma_{J_{k}}^{-1}x_{n+1,J_{k}}\right)\\
  =&c' \gamma_j(k)\,,
\end{align*}
for some $c,c'>0$.
Therefore $\eta_{j}(k+1)$ as the same sign as $\gamma_j(k)$.
But $v_j(t_{k+1})=v_j(t_{k})+\gamma_j(k)(t_{k+1}-t_{k})$, thus
$v_j(t_{k+1})$ must have the same sign as $\gamma_j(k)$.
As a result, we have confirmed that $\eta_j(k+1)$ and $v_j(t_{k+1})$
have the same sign.
\end{proof}

\subsection{The conformal Lasso algorithm}
Now we describe an efficient and exact algorithm to compute
the Lasso conformal prediction set.  We introduce a few more notations.
\begin{itemize}
  \item $r_i(t)=y_i-x_i'\hat\beta(t)$ ($i=1,...,n,n+1$) denotes the fitted residual
  when using $(x_{n+1},y_{n+1}(t))$ as the $(n+1)$th data point, where
  $y_{n+1}(t)=x_{n+1}'\hat\beta +t$.
  \item Throughout the rest of this paper we let $\hat C(x_{n+1})$ be the
  Lasso conformal prediction set at $x_{n+1}$, using the negative absolute
  fitted residual \eqref{eq:conformity} as conformity score function.
  \item For a set $C$ and a number $z$, $C+z=\{y: y-z\in C\}$ denotes the shifted set. 
\end{itemize}

By definition, $y_{n+1}(t)\in\hat C(x_{n+1})$ if and only if
the rank of $|r_{n+1}(t)|$ among all fitted residuals (including itself)
is no higher than $(n+1)\alpha$ in decreasing order.

\Cref{thm:continuation} implies that the Lasso solution
$\hat\beta(t)$ is piecewise linear.  Therefore,
the fitted residuals $r_{i}(t)$ are also linear in $t$ between
a pair of consecutive points of change $(t_k,t_{k+1})$.
Then one can easily find all values of $t$ in $(t_k,t_{k+1})$
at which the rank of $|r_{n+1}(t)|$ changes, by solving 
$|r_{n+1}(t)|=|r_i(t)|$ for $t$ in the interval $[t_k,t_{k+1}]$.
Denote these points by $t_{k}=t_{k,0}< t_{k,1}<t_{k,2}<...<t_{k,\ell_k}=t_{k+1}$.
Between these points, the rank of $|r_{n+1}(t)|$ is constant.
Then we have
\begin{equation}\label{eq:conf_segment}
  \left[\hat C(x_{n+1})-y_{n+1}(0)\right]\cap [t_k,t_{k+1})=
  \bigcup_{\ell\in L_k}[t_{k,\ell},t_{k,\ell+1})
\end{equation}
where $L_k$ consists of all $\ell$'s in $\{0,...,\ell_k-1\}$
such that the rank of $|r_{n+1}(t_{k,\ell})|$ among $\{|r_i(t_{k,\ell})|:1\le i\le n+1\}$
is no higher than $(n+1)\alpha$ in decreasing order.

We summarize the algorithm of conformal prediction using Lasso in \Cref{alg:conf_lasso}.
Some of the steps, such as finding the rank
of fitted residuals and solving linear equations, are simple enough 
so the details are omitted.
\begin{algorithm}[tb]
\caption{Conformal Prediction with Lasso}
\label{alg:conf_lasso}
\begin{algorithmic}
\State{\bf Input:} Data $(x_i,y_i)_{i=1}^n$, new covariate
$x_{n+1}$, range $(y_{\min},y_{\max})$, miscoverage
  level $\alpha \in (0,1)$, Lasso parameter $\lambda$ 
\State{\bf Output:} $\hat D=\hat C(x_{n+1})\cap [y_{\min},y_{\max}]$.
\\\hrulefill
\State Let $\hat\beta(0)$ be the Lasso solution on data $(x_i,y_i)_{i=1}^n$ with tuning parameter $\lambda$
\State $y_{n+1}(0)\leftarrow x_{n+1}'\hat\beta(0)$, $t_0\leftarrow 0$, $k\leftarrow 0$
\State $\hat D_+\leftarrow\{y_{n+1}(0)\}$
\While{$t_0+...+t_k<y_{\max}-y_{n+1}(0)$}
  \State Let  $J_k$ be the active set of $\hat\beta(t_k)$.
  \State Calculate $\eta(k)$, $\gamma(k)$, $t_{k+1}$ as in \eqref{eq:rec_eta+1}, \eqref{eq:rec_gamma+1}, 
  \eqref{eq:t_k+1}.
  \State $t_{k+1}\leftarrow \min(t_{k+1},y_{\max}-y_{n+1}(0)-(t_0+...+t_k))$.
  \State $\{t_{k,1},...,t_{k,\ell_k-1}\}\gets\{t\in(t_k,t_{k+1}):|r_{i}(t)|=|r_{n+1}(t)|~\text{for some}~1\le i\le n\}$.
  \State $L_k\gets\left\{1\le\ell< \ell_k:\sum_{i=1}^{n+1}\mathbf 1\left[|r_i(t_{k,\ell})|\le |r_{n+1}(t_{k,\ell})|\right]\le \lceil(n+1)(1-\alpha)\rceil\right\}$ 
  \State $\hat D_+\leftarrow \hat D_+\bigcup \left\{\cup_{\ell\in L_k}[t_{k,\ell},t_{k,\ell+1})\right\}$.
  \State $k\leftarrow k+1$
\EndWhile
\State Repeat the above procedure analogously for negative values of $t$, obtaining $\hat D_-$.
\State Return $\hat D=y_{n+1}(0)+(\hat D_+\cup \hat D_-)$.
\end{algorithmic}
\end{algorithm}

\begin{remark}
  \Cref{alg:conf_lasso} requires an interval $[y_{\min},y_{\max}]$
  as part of the input.  Theoretically speaking, this interval
  can be chosen simply as $[y_{(1)},y_{(n)}]$, the sample range of the response variable,
  where $y_{(1)}\le y_{(2)}\le \ldots \le y_{(n)}$ are the order statistics of the response variable.
  Doing this will incur a loss of coverage no more than $2/(n+1)$, because
  $P(Y_{n+1}\in [Y_{(1)},Y_{(n)}])\ge 1-2/(n+1)$.
  In our numerical experiments, we set the search range even more conservatively,
  enlarging the sample range by $50\%$
  of length
  \begin{equation}\label{eq:range}
  [y_{\min},y_{\max}]=\left[y_{(0)}-0.25(y_{(n)}-y_{(0)}),~~y_{(n)}+0.25(y_{(n)}-y_{(0)})\right]\,.
  \end{equation}
\end{remark}

\subsection{When is the Lasso conformal prediction set an interval?}
Being an interval is a conceptually and practically 
desirable property of a prediction set.  From the original definition, it is
generally unclear if the Lasso conformal prediction set is an interval
at a particular covariate value $x_{n+1}$.  
\Cref{thm:continuation} and \Cref{rem:simple} imply
that the conformal prediction set must be a union
of intervals.  However, as we explain next, very often the prediction set is an 
interval.  

Consider $t\in[t_k,t_{k+1}]$ for a pair of consecutive points of change $(t_{k},t_{k+1})$.
We focus on the case $t_{k}>0$.
For $1\le i\le n$, the residuals $r_i(t)=y_i-x_i'\hat\beta(t)$ is a linear function
satisfying 
\begin{align}
  r_i(t)=&r_i(t_k)-x_{i,J_k}'\eta(k)(t-t_k)\nonumber\\
  =&r_i(t_k)-\frac{n^{-1}x_{i,J_k}'\hat\Sigma_{J_k}^{-1}x_{n+1,J_k}}{1+n^{-1}x_{n+1,J_k}'
  \hat\Sigma_{J_k}^{-1}x_{n+1,J_k}}(t-t_k)\,,\label{eq:r_i(t)}
\end{align}
whereas for the $(n+1)$th data point
\begin{align}
r_{n+1}(t)=&r_{n+1}(t_k)+(t-t_k) - x_{n+1,J_k}'\eta(k)(t-t_k)\nonumber\\
=&r_{n+1}(t_k)+\frac{1}{1+n^{-1}x_{n+1,J_k}'\hat\Sigma_{J_k}^{-1}x_{n+1,J_k}}(t-t_k)\,.\label{eq:r_n+1_(t)}
\end{align}

According to \eqref{eq:r_n+1_(t)}, when $t$ moves away from $0$, the fitted residual $r_{n+1}(t)$
does not change its sign and its absolute value is strictly increasing in a piecewise
linear manner with a slope of $(1+n^{-1}x_{n+1,J_k}'\hat\Sigma_{J_k}^{-1}x_{n+1,J_k})^{-1}$ in the $k$th piece
of the homotopy,
which is close to $1$ if $n$ is large and $\hat\Sigma_{J_k}$ is well-conditioned.
Similarly, \eqref{eq:r_i(t)} suggests that when $\hat\Sigma_{J_k}$ is well-conditioned and $n$
large, the residuals $r_i(t)$ change slowly for $1\le i\le n$.

Combining these two observations, we expect that in most cases, $r_{n+1}(t)$ changes
faster than $r_i(t)$ ($1\le i\le n$).  So $|r_i(t)|=|r_{n+1}(t)|$ only happens
for one negative value and one positive value of $t$.  As a result, the rank of
$|r_{n+1}(t)|$ increases monotonically as $t$ moves away from $0$.  
In this case, $\hat C(x_{n+1})$
is an interval containing $y_{n+1}(0)$.  Formally we have the following result.
\begin{theorem}\label{thm:conf_int}
  If
  \begin{align}\label{eq:conf_int_condition}
     \max_{1\le i\le n}\left|x_{i,J}'\hat\Sigma_J^{-1}x_{n+1,J}\right| < n\,,
  \end{align}
  for all the active sets $J$ 
  as $t$ varies over the real line, then
  the Lasso conformal prediction set is an interval.
\end{theorem}
\begin{remark}
  The quantity $x_{i,J}'\hat\Sigma_{J} x_{i,J}$ is the leverage score
  of the $i$th sample point constrained on the active set $J$.
  In ordinary least square regression, a small leverage score implies
  robustness at $x_{i}$.  In our analysis, it turns out that
  $x_{i,J}'\hat\Sigma_{J} x_{n+1,J}$, $1\le i\le n$, which we call the \emph{cross leverage scores},
  measure the stability of fitted residuals under the addition of a new sample point
  at $x_{n+1}$ when the active set is $J$. Further discussion about robustness
  and stability is given in \Cref{sec:discussion}.
\end{remark}

\paragraph{Efficient implementation of \Cref{alg:conf_lasso}.} Although the condition of
 \Cref{thm:conf_int} is hard to verify over all values of $t$.  It is easy
 to verify for any given pair of consecutive points of change.
 Suppose we have found $0\le t_k<t_{k+1}$ and need to find $\left[\hat C(x_{n+1})-y_{n+1}(0)\right]\cap [t_k,t_{k+1}]$
 in \Cref{alg:conf_lasso}.  If \eqref{eq:conf_int_condition}
 holds for $J=J_k$, then we have
 \begin{align*}
   &\left[\hat C(x_{n+1})-y_{n+1}(0)\right]\cap [t_k,t_{k+1})\\
   =&\left\{
   \begin{array}{ll}
     \emptyset, & \text{if}~~ t_{k} \notin \hat C(x_{n+1})\,,\\
     \mbox{$[t_k,t_{k+1})$}, & \text{if}~~ t_{k} \in \hat C(x_{n+1}),~~t_{k+1}\in \hat C(x_{n+1})\,,\\
      \mbox{$[t_k, t^*)$}, & \text{if}~~t_{k}\in \hat C(x_{n+1}),~~t_{k+1}\notin\hat C(x_{n+1})\,,
   \end{array}\right.
 \end{align*}
 where $t^*\in (t_k,t_{k+1})$ is the unique value that satisfies $|r_i(t^*)|=|r_{n+1}(t^*)|$ for some $1\le i\le n$ and the
 rank of $|r_{n+1}(t)|$ crosses the threshold $\lceil (n+1)(1-\alpha)\rceil$ at $t^*$.
This leads to a substantial speedup 
 of \Cref{alg:conf_lasso}.

\paragraph{An accuracy-stability trade-off.} According to \eqref{eq:r_i(t)} and \eqref{eq:r_n+1_(t)},
if $n^{-1}x_{i,J_k}\hat\Sigma_{J_k}^{-1}x_{n+1,J_k}$ is much smaller than $1$ for all
$i$ and all active sets $J_k$ on the solution path, then $|r_{n+1}(t)|$ will quickly
become larger than other $|r_{i}(t)|$'s as $t$ moves away from $0$.  Thus, for larger values of $\lambda$,
the size of $J_k$ tend to be smaller and  $x_{i,J_k}\hat\Sigma_{J_k}^{-1}x_{n+1,J_k}$
is also smaller,
which leads to a shorter conformal prediction interval.  We call this the benefit
of stability: The original fitted residuals $r_i(t)$ changes slowly under the perturbation of the
$(n+1)$th data point with a slope near $0$, while the $(n+1)$th fitted residual $r_{n+1}(t)$ changes
much faster with a slope near $1$.
 On the other hand, when $\lambda$ is large,
the Lasso algorithm searches for $\hat\beta$ over a smaller feasible set, 
thus the estimate suffers from an inferior accuracy:
The fitted residuals $r_i(0)$ may be large to start with, which leads to a wider
conformal prediction interval.  Therefore we have the accuracy-stability trade-off in
conformal prediction: $\lambda$ needs to be large enough for the
fitted residuals to be stable, but not too large so the fitted residuals are small.
 
\subsection{Extension to the elastic net}\label{sec:en}
The elastic net \citep{enet} adds an $\ell_2$ penalty to the Lasso problem.  The resulting
objective function is then strongly convex and the solution is more stable.
Our derivation for the Lasso easily extends to the elastic net.
Now we consider the problem
\begin{align}
  \hat\beta^{\rm (en)}(t)=&
  \arg\min_{\beta}\Bigg\{\frac{1}{2}\sum_{i=1}^n(y_i-x_i'\beta)^2\nonumber\\
  &\quad+\frac{1}{2}
  \left[x_{n+1}'\hat\beta^{\rm (en)}(0)+t-x_{n+1}'\beta\right]^2+
  \lambda\|\beta\|_1+\frac{\rho}{2}\|\beta\|_2^2\Bigg\}\,,\label{eq:en}
\end{align}
where $$\hat\beta^{\rm (en)}(0)=
  \arg\min_{\beta}\frac{1}{2}\sum_{i=1}^n(y_i-x_i'\beta)^2
  +\lambda\|\beta\|_1+\frac{\rho}{2}\|\beta\|_2^2$$ is the elastic net estimate 
  using the original data.  A similar homotopy holds for the
  elastic net solution path indexed by $t$, using an almost identical derivation.

\begin{corollary}\label{cor:en_homotopy}
  \Cref{lem:beta_0} and \Cref{thm:continuation}
  hold for the solution
  path $\{\hat\beta^{\rm (en)}(t):t\in\mathbb R\}$, with
  $n\hat\Sigma_{J_k}$ replaced by $n\hat\Sigma_{J_k}+\rho I_{|J_k|}$ for all
  $k\ge 0$, where
  $I_m$ denotes the $m\times m$ identity matrix.
\end{corollary}

Using the analogous versions of \eqref{eq:r_i(t)} and \eqref{eq:r_n+1_(t)}
with with
  $n\hat\Sigma_{J_k}$ replaced by $n\hat\Sigma_{J_k}+\rho I_{|J_k|}$, we
  obtain an easy to verify sufficient condition for the elastic net
  conformal prediction set to be an interval.
\begin{corollary}\label{cor:en_int}
  If $\rho\ge \|x_{n+1}\|\cdot\max_{1\le i\le n}\|x_i\| $, then
  the elastic net conformal prediction set at $x_{n+1}$
  is an interval.
\end{corollary}

\section{Numerical experiments}
\label{sec:data}

We examine the statistical and numerical performance of the conformal Lasso
algorithm over a wide range of synthetic data sets and two real data examples.

\subsection{Synthetic data examples}
Our synthetic data examples follow the setup of those in \cite{LeiGRTW16}.
We consider both low dimensional and high dimensional settings.

\paragraph{Low dimensional setting.} In the low dimensional setting, the sample size is set to be $n=100$,
and the covariate has dimensionality $p=10$.
We consider three different models to generate the data pair $(X,Y)$.
\begin{itemize}
  \item Model I (Standard Gaussian linear model). In this model
  $Y=X'\beta+\epsilon$, where $\beta$ is a $p$-dimensional vector
  whose entries are $1$ or $-1$ with a random sign. The marginal distribution
  of $X$ is $N(0,I_p)$ where $I_p$ is the identity matrix.  The noise $\epsilon$
  has a standard normal distribution, and is independent of $X$ and $\beta$.
  \item Model II (Nonlinear additive model). This model uses
  $Y=\sum_{j=1}^p f_j(X(j))+\epsilon$, where each $f_j$ is a B-Spline
  function with $4$ degrees of freedom (using function \texttt{bs} in \texttt{R} 
  package \texttt{splines}) and coefficients randomly generated
  from $\{-1,1\}$.  Again, $\epsilon$ is standard normal and independent of $X$
  and all $f_j$'s.
  \item Model III (linear model with Non-Gaussian correlated design and heavy-tailed noise).
  The model is linear: $Y=X'\beta+\epsilon$ with $\beta$ generated the same way as in 
  Model I.  The covariate matrix $X$ is generated by taking
  a column-wise weighted moving average of a random matrix $Z$ using iid ${\rm Uniform(0,1)}$
  weights.  Each column of $Z$ is randomly sampled from one of the three
  distributions: normal, Bernoulli with parameter $0.5$, skewed normal with skewness
  parameter $5$.  The columns of $Z$ are scaled to have the same variance.
  Finally, the noise has a student $t$-distribution with $2$ degrees of freedom.
  We note that the noise, and hence the response variable $Y$,
  does not have a finite second moment.
\end{itemize}

For each model, we generate a data set of $n=100$ training sample points, 
each accompanied with $100$ testing sample points.  A conformal prediction set
with target coverage level $0.9$ ($\alpha=0.1$)
is calculated for each testing sample point using each of the three methods:
the exact conformal lasso algorithm presented in this paper; the grid point
evaluation method implemented in the \texttt{conformalInference} package \citep{LeiGRTW16};
the split conformal method proposed in \citep{LeiGRTW16}.  The base estimator is
Lasso with $\lambda$ chosen by the median of the cross-validated values from $100$
independent samples of the same size.
The search range used in both the exact method and the grid evaluation method
is chosen according to \eqref{eq:range}.

The experiment is repeated on
$100$ independently generated samples.  We report the average empirical coverage, average size of the conformal
prediction set (length of interval), and average running time per data set (in seconds) in
  \Cref{tab:sim_low_1,tab:sim_low_2,tab:sim_low_3} 
 for models I, II, III, respectively.  One standard error is given in the parenthesis
 following the average number.

From the simulation we observe that all three methods provide valid and nearly perfect
coverage.  The grid method and exact method give similar lengths, where the slight
difference is due to the rounding between neighboring grid points in the grid method.
In this setting, the exact method is much faster than the grid method, with same solid
performance.

\begin{table}[h]
  \begin{center}
\begin{tabular}{|l|l|l|l|}
\hline
& grid & split & exact \\
\hline
Coverage & $0.911~(0.004)$  & $0.910~(0.004)$ & $0.905~(0.004)$  \\
\hline
Length & $3.57~(0.03)$ & $3.77~(0.05)$  & $3.51~(0.03)$   \\
\hline
Time & $32.2~(0.07)$ & $0.008~(<10^{-3})$ & $0.045~(<10^{-3})$   \\
\hline
\end{tabular}
\caption{Simulation 1(I): Low-dimensional setting with the standard linear model.
\label{tab:sim_low_1}}
\end{center}
\end{table}

\begin{table}[h]
  \begin{center}
\begin{tabular}{|l|l|l|l|}
\hline
& grid & split & exact \\
\hline
Coverage & $0.897~(0.005)$  & $0.890~(0.006)$ & $0.898~(0.005)$  \\
\hline
Length & $5.95~(0.13)$ & $7.29~(0.30)$  & $5.98~(0.14)$   \\
\hline
Time & $30.5~(0.03)$ & $0.008~(<10^{-3})$ & $0.076~(0.002)$   \\
\hline
\end{tabular}
\caption{Simulation 1(II): Low-dimensional setting with the nonlinear additive model.
\label{tab:sim_low_2}}
\end{center}
\end{table}

\begin{table}[h]
  \begin{center}
\begin{tabular}{|l|l|l|l|}
\hline
& grid & split & exact \\
\hline
Coverage & $0.909~(0.004)$  & $0.906~(0.005)$ & $0.905~(0.004)$  \\
\hline
Length & $18.0~(0.53)$ & $20.1~(0.95)$  & $17.5~(0.50)$   \\
\hline
Time & $32.3~(0.07)$ & $0.008~(<10^{-3})$ & $0.070~(0.002)$   \\
\hline
\end{tabular}
\caption{Simulation 1(III): Low-dimensional setting with the non-Gaussian, correlated, heavy-tail, linear model.
\label{tab:sim_low_3}}
\end{center}
\end{table}

\paragraph{High dimensional setting.}
In the high dimensional setting, the sample size is $n=200$, and the
covariate dimensionality is $p=500$.
Similarly we consider three models, but with sparsity.
\begin{itemize}
  \item Model I (sparse Gaussian linear model).  The difference
  from the low dimensional case is that only
  the first $5$ coordinates of $\beta$ are non-zero, and set to be
  $-8$ or $8$ with signs chosen at random.
  \item Model II (sparse nonlinear additive model). The difference from the low dimensional
  setting is that only $5$ $f_j$'s are non-zero, and the 
  linear coefficients of the spline bases are set to be $-8$ or $8$
  with signs chosen at random.
  \item Model III (non-Gaussian, correlated, heavy-tail, linear model).  The difference
  is that only first $5$ coordinate of $\beta$ are non-zero, and set to be $-8$ or $8$, with
  signs chosen at random.
 \end{itemize}
 
 The simulation is carried out in the same manner as in the low dimensional setting.
 For computational efficiency, instead of searching the entire interval
 specified by \eqref{eq:range}, the exact algorithm stops when it finds the end points
 of the interval in the conformal prediction set that contains $x_{n+1}'\hat\beta$.
 The results are summarized in \Cref{tab:sim_hi_1,tab:sim_hi_2,tab:sim_hi_3}.
 
 A notable observation in the high dimensional setting is a substantial
 over-conservative coverage for the grid method in Model I.  This is
a rounding error due to an \textit{ad hoc} interpolation between two neighboring grid points
 used by the algorithm when finding the end points of the prediction interval.
 Such a rounding error becomes substantial when the grid is not dense enough but
 the response variable happens to have non-negligible probability mass near the cut-off point.
 This illustrates an advantage of the exact method, as it offers exactly
 the desired level of coverage, often with a shorter interval.
 Again, the computing time of the exact method still compares favorably against
 the grid method, even when the grid is sparse. 
 
 \begin{table}[h]
   \begin{center}
 \begin{tabular}{|l|l|l|l|}
 \hline
 & grid & split & exact \\
 \hline
 Coverage & $0.937~(0.003)$  & $0.899~(0.004)$ & $0.895~(0.004)$  \\
 \hline
 Length & $4.15~(0.02)$ & $4.09~(0.05)$  & $3.61~(0.02)$   \\
 \hline
 Time & $93.9~(0.6)$ & $0.01~(<10^{-3})$ & $7.8~(0.07)$   \\
 \hline
 \end{tabular}
 \caption{Simulation 2-I: High-dimensional setting with the standard linear model.
 \label{tab:sim_hi_1}}
 \end{center}
 \end{table}

 \begin{table}[h]
   \begin{center}
 \begin{tabular}{|l|l|l|l|}
 \hline
 & grid & split & exact \\
 \hline
 Coverage & $0.905~(0.004)$  & $0.900~(0.005)$ & $0.903~(0.004)$  \\
 \hline
 Length & $23.55~(0.85)$ & $30.5~(1.85)$  & $23.46~(0.85)$   \\
 \hline
 Time & $50.4~(0.85)$ & $0.01~(<10^{-3})$ & $10.2~(0.87)$   \\
 \hline
 \end{tabular}
 \caption{Simulation 2-II: Hi-dimensional setting with the nonlinear additive model.
 \label{tab:sim_hi_2}}
 \end{center}
 \end{table}

 \begin{table}[h]
   \begin{center}
 \begin{tabular}{|l|l|l|l|}
 \hline
 & grid & split & exact \\
 \hline
 Coverage & $0.902~(0.004)$  & $0.904~(0.004)$ & $0.897~(0.004)$  \\
 \hline
 Length & $15.9~(0.63)$ & $17.9~(1.39)$  & $15.2~(0.59)$   \\
 \hline
 Time & $44.8~(0.19)$ & $0.010~(<10^{-3})$ & $5.64~(0.01)$   \\
 \hline
 \end{tabular}
 \caption{Simulation 2-III: High-dimensional setting with the non-Gaussian, correlated, heavy-tail, linear model.
 \label{tab:sim_hi_3}}
 \end{center}
 \end{table}
 
 \subsection{Real data examples}
 We apply the exact conformal Lasso algorithm to two data sets: the
 diabetes data \citep{lars} and the Boston housing data \citep{Harrison78}.
 
 \paragraph{The diabetes data.}
 The diabetes data is considered in \cite{lars} to illustrate basic properties of
 the LARS algorithm.  It contains $442$ subjects.
 Each subject has ten covariates, including age, gender, body mass index, blood pressure,
 and six blood serum measurements.  The response is a continuous measurement of
 diabetes progression one year after the initial measurement.
 According to \cite{lars}, one may either consider the linear model that uses a regularized
 linear function of the ten covariates to predict the response, or one may fit
 a regularized quadratic function as a linear combination of the 10 original covariates 
 and 54 second order
 terms.  For ease of interpretation, we center and scale the response variable so that
 the sample variance is $1$.
 
 In our data examples, we implement the algorithm with a more realistic way
 of choosing the Lasso tuning parameter.  To this end, we randomly split the data set into
 a fitting subsample of size $300$, and a testing sample of size $142$.
 We use the cross-validation with confidence method \citep[CVC,][]{Lei17}
 on the fitting subsample
 to choose a value of $\hat\lambda$ that gives a parsimonious model fit with
 competitive predictive risk. Then we consider four values of $\lambda$:
 $(2^{1/2},1,2^{-1/2},2^{-1})\hat\lambda$.  For each value of $\lambda$ we construct
 the Lasso conformal prediction set at target coverage level $0.9$ for each data
 point in the testing sample using
 the search range given in \eqref{eq:range}.
 The procedure is repeated with $100$ independent sample splittings.
 We report four quantities: the average coverage, average length of the prediction set,
 average number of linear pieces on the homotopy covered in the search, and average number of active variables
 on the linear homotopy covered in the search.

\begin{figure}[t]
   \begin{center}
     \includegraphics[scale=\myscale]{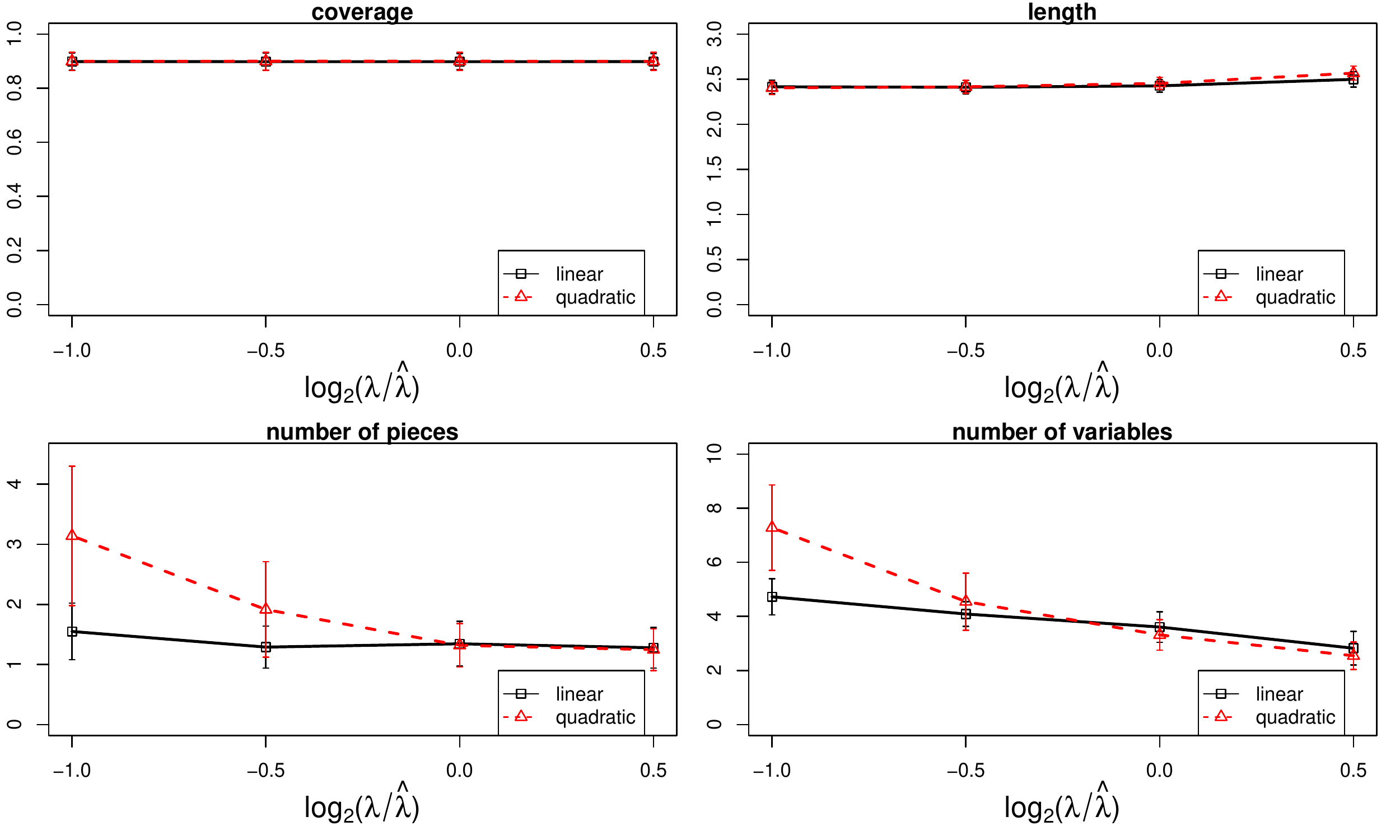}
     \caption{Results for the diabetes data. The $x$-axis indexes
     four values of $\lambda$, where $\hat\lambda$ is the
     tuning parameter chosen by the CVC method.\label{fig:diabetes}}
   \end{center}
 \end{figure}

 The results are summarized in \Cref{fig:diabetes}.  The averaged quantities of interest
 are plotted as a function of $\log_2(\lambda/\hat\lambda)$, together with one standard
 deviation.
 Here the choice of $\lambda$ has very little effect on the coverage,
 as all four values of $\lambda$ give the right empirical coverage.
 We also observe that the tuning parameter value exhibits a trade-off
 between the prediction accuracy (size of prediction interval),
 computational efficiency (number of linear pieces on the homotopy path),
 and interpretability (number of active variables on the homotopy path).
 In particular, the choice $\lambda=\hat\lambda$ given by the CVC method reaches
a good balance in this trade-off.
 Moreover, the linear model seems to have the same predictive power when compared
 to the quadratic model, while being more computationally efficient.
 
 \paragraph{The Boston housing data.}
 The Boston housing data consists of demographic and housing condition information 
 of $506$ areas in Boston \citep{Harrison78}.  There are $13$ covariates including
 demographic information, location, air quality, and housing condition.
 The response variable is the median value of owner-occupied homes.  All variables
 are centered and standardized to have unit variance.
 
 \begin{figure}[t]
   \begin{center}
     \includegraphics[scale=\myscale]{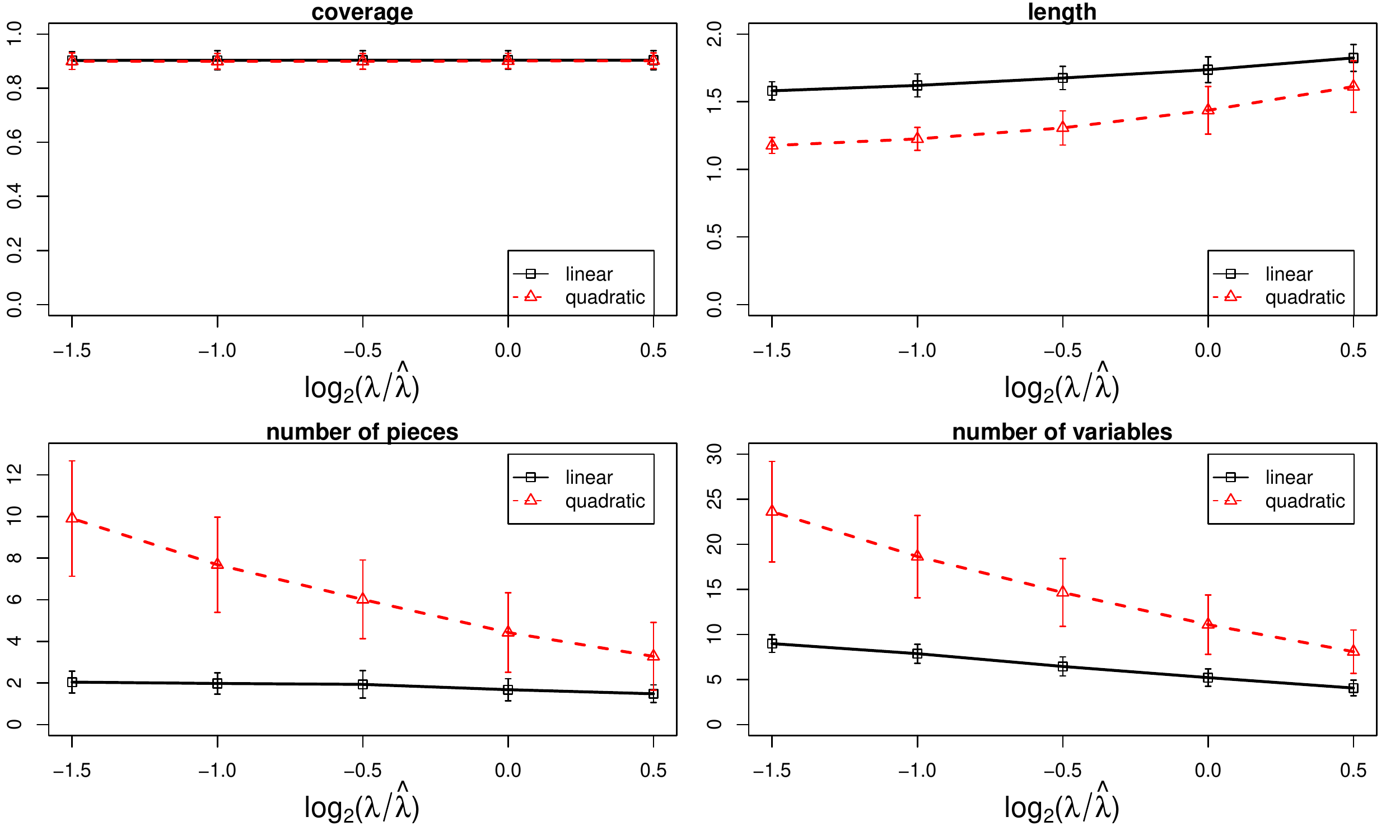}
     \caption{Results for the Boston housing data. The $x$-axis indexes
     five values of $\lambda$, where $\hat\lambda$ is the
     tuning parameter chosen by the CVC method.\label{fig:boston}}
   \end{center}
 \end{figure}
 
 The numerical experiment is conducted in a similar way as in the diabetes data,
 including both the linear model and quadratic model with $103$ covariates.
 We randomly split the data into a fitting sample of size $400$, and a testing
 sample of size $106$.  The results for $100$ independent sample splits are plotted in \Cref{fig:boston}.
 
 A similar observation is that all values of $\lambda$ give the right level of average
 coverage with very small variability.
 There are also two notable differences from the diabetes data.
 First, the Boston housing data exhibits a higher degree of predictability than
 the diabetes data. Comparing the upper right panels of \Cref{fig:boston,fig:diabetes},
  the $90\%$ Lasso conformal prediction intervals in the Boston housing data
 are much shorter than those in the diabetes data with comparable levels of
 regularization.
 Second, the inclusion of quadratic terms significantly improves prediction for
 the Boston housing data, with a substantial drop in the length of
 prediction intervals.  On the contrary, the linear and quadratic model give
 nearly the same prediction interval lengths in the
 diabetes data.
 Finally, in Boston housing data, smaller values of $\lambda$ tend to produce 
 shorter prediction intervals, using a moderately large number of variables, 
 which suggests that the covariates carry rich information about the response 
 and little regularization is needed.

\section{Discussion}
\label{sec:discussion}
Conformal prediction has deep connections to statistical techniques such as
jackknife, bootstrap, and cross-validation, in the use of estimates obtained
from slightly perturbed data.  It is also linked to the notion of algorithmic 
stability, which has gained substantial attention in both the statistics and machine learning
literature, and has been
proved fundamentally important in both
learning theory and statistical reproducibility
\citep{SSSS10,Yu13,WangLF16}. 
Roughly speaking, algorithmic stability requires the estimate to change little 
when one data
entry is changed arbitrarily.  
Conformal prediction reflects
the importance of algorithmic stability from a slightly different perspective, 
which we call \emph{residual
stability}.  When the residuals are
stable, then the conformalized residuals $(r_i:1\le i\le n)$ 
will change slowly when the trial response
value $y_{n+1}$ changes, while the last residual $r_{n+1}$ changes almost
linearly with a slope close to one.  Then by construction, the conformal prediction set
is a narrow interval, as $|r_{n+1}|$ quickly outgrows $(|r_i|:1\le i\le n)$
as $y_{n+1}$ moves away from the initial predicted value $\hat y_{n+1}=x_{n+1}'\hat\beta$. 

In the case of Lasso, although it is known that the support recovery cannot
be algorithmically stable \citep{NoFreeLunch}, the fit is usually stable
when $\lambda$ is conventionally chosen. Our experiments have confirmed
this in many different settings using both synthetic and real data, where
\eqref{eq:conf_int_condition} holds even when the active set changes frequently
on the homotopy path.  Understanding the residual stability of
prediction algorithms would be an important topic for future research.

The current paper provides an initial understanding of the residual stability
of the Lasso, which leads to a simple and efficient algorithm of conformalizing
 the Lasso estimator.  It is possible to extend the methods and results
to other settings, such as generalized Lasso \citep{TibsT11}, graphical Lasso \citep{glasso}, and sparse subspace
estimation \citep{Jolliffe03spca,dAspremont05direct,Zou06spca,Vu13,ChenL15,LeiV15}.


%


\bibliographystyle{asa}
\bibliography{paper}
\end{document}